\documentclass[11pt,english]{article}
\usepackage{palatino}
\usepackage[T1]{fontenc}
\usepackage[latin1]{inputenc}
\usepackage{geometry}
\usepackage{calc}
\usepackage{amsmath}
\usepackage{graphicx}
\usepackage{amssymb}

\makeatletter

\newcommand{\lyxaddress}[1]{
\par {\raggedright #1
\vspace{1.4em}
\noindent\par}
}

\usepackage{babel}
\makeatother
\begin{document}

\title{Self-Organization and the Selection of Pinwheel Density in Visual
Cortical Development\date{} }

\author{Matthias Kaschube$^{1-4}$, Michael Schnabel$^{1,2}$ and Fred Wolf$^{1,2}$}

\maketitle

\lyxaddress{$^{1}$Max Planck Institute for Dynamics and Self-Organization and
$^{2}$Bernstein Center for Computational Neuroscience, Goettingen,
Germany, $^{3}$Lewis-Sigler Institute for Integrative Genomics and
$^{4}$Joseph Henry Laboratories of Physics, Princeton University,
Princeton NJ, USA}

\textbf{Self-organization of neural circuitry is an appealing framework
for understanding cortical development, yet its applicability remains
unconfirmed. Models for the self-organization of neural circuits have
been proposed, but experimentally testable predictions of these models
have been less clear. The visual cortex contains a large number of
topological point defects, called pinwheels, which are detectable
in experiments and therefore in principle well suited for testing
predictions of self-organization empirically. Here, we analytically
calculate the density of pinwheels predicted by a pattern formation
model of visual cortical development. An important factor controlling
the density of pinwheels in this model appears to be the presence
of non-local long-range interactions, a property which distinguishes
cortical circuits from many nonliving systems in which self-organization
has been studied. We show that in the limit where the range of these
interactions is infinite, the average pinwheel density converges to
$\pi$. Moreover, an average pinwheel density close to this value
is robustly selected even for intermediate interaction ranges, a regime
arguably covering interaction-ranges in a wide range of different
species. In conclusion, our paper provides the first direct theoretical
demonstration and analysis of pinwheel density selection in models
of cortical self-organization and suggests to quantitatively probe
this type of prediction in future high-precision experiments.}

\section{Introduction}

Neuronal circuits in the mammalian cerebral cortex are among the most
complex systems in nature. The biological mechanisms that contribute
to their formation in early brain development remain poorly understood.
However, it is unlikely that the precise architecture of mature cortical
circuits can be attributed to genetic prespecification, since the
number of genes in the genome is insufficient\cite{spitzer:06}. Instead,
dynamical self-organization presumably plays a major role in shaping
the architecture of neuronal circuits in the cerebral cortex. Dynamical
self-organization is most thoroughly described in non-living physical
systems driven outside of thermodynamic equilibrium by external forcing.
Whereas the emergence of structure is externally driven, the structures
formed are primarily determined through interactions within the system
itself\cite{cross:93,ball:99}. Neural circuits encompass various
positive and negative feedback loops, and these could well form the
basis for cortical self-organization. However, evidence supporting
this presumption is derived from theoretical considerations\cite{shatz:90,singer:95,ooyen:03}
rather than empirical observation. Models for the self-organization
of neural circuits have been proposed, but experimentally testable
predictions of these models have been lacking. 

The system of orientation columns in the visual cortex is a paradigmatic
system for studying cortical development and the role of self-organization
in this process. Most neurons in the visual cortex respond selectively
to a particular orientation of an elongated visual stimulus. Whereas
in columns perpendicular to the cortical surface, neurons prefer similar
stimulus orientations, the preferred orientation varies mostly smoothly\cite{hubel:62}
and repetitively across the cortical surface giving rise to a complex
two-dimensional pattern called the map of orientation preference (Fig.
1a, b). Throughout the cortical map, there are point-like orientation
singularities\cite{blasdel:86,bonhoeffer:91,bonhoeffer:93} called
pinwheel centers\cite{braitenberg:79} at which all stimulus orientations
are represented in circular fashion. Numerous studies are consistent
with the hypothesis that orientation maps develop through activity-dependent
self-organization. They form in dark reared animals\cite{white:01},
under substantial manipulation of visual input\cite{crair:98,sengpiel:99},
and even in auditory cortex when rewired to be driven by visual inputs\cite{sharma:00}.
Moreover, an analogy between cortical development and pattern formation
appears plausible. Like in other systems where pattern formation has
been observed \cite{cross:93,ball:99}, the orientation map arises
probably from an initially non-selective state, it exhibits a typical
periodicity and a spatial extension at least an order of magnitude
larger than the basic periodicity length.

The conditions under which orientation maps can arise through self-organization
have been thoroughly investigated theoretically\cite{miller:94,swindale:96,wolf:98}.
A recent and highly promising approach stressing the analogy to pattern
forming systems showed that pinwheels can be stabilized by activity-dependent
long-range interactions \cite{wolf:05}. A phenomenological order
parameter field model based on the Swift-Hohenberg equation \cite{swift:77}
was proposed in which orientation maps arise from a supercritical
bifurcation of Turing-type. In this class of models, the stabilizing
nonlinearity includes only key features of visual cortical organization
and is constraint by biologically plausible symmetry assumptions.
The model exhibits multiple structurally distinct quasiperiodic attractors
resembling orientation maps in the visual cortex. 

The qualitative similarity of solutions of this model to orientation
maps in the visual cortex appears promising. However, it is unclear
at present weather the model accounts for aspects of cortical organization
also quantitatively and whether this resemblance is not just superficial.
Comparing directly orientation maps in the model and the visual cortex
would be difficult, because of the large number of possible map structures.
Instead, it will be virtually unavoidable to take a statistical approach
to this questions. As discrete entities, pinwheels can be detected,
counted and localized with high accuracy. Here, we calculate in the
model the average density of pinwheels in the limit of infinite interaction
range. We show that this density is representative for a large regime
of intermediate interaction ranges covering the estimated ranges in
various mammalian species. Therefore, the quantity pinwheel density
appears particularly well suited for testing for signatures of long-range
dominated self-organization in experiment. In the following, we briefly
describe the system of orientation preferences in the visual cortex
and present the long-range self-organization model for its activity-dependent
development.

\subsection{Orientation preference maps}

\begin{figure}
\begin{centering}\includegraphics[scale=0.7]{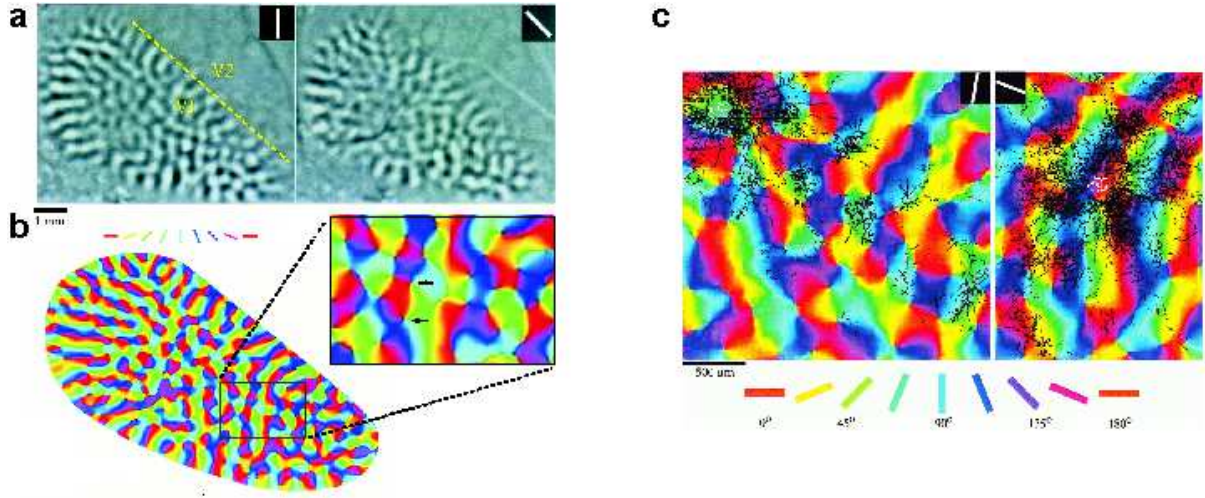}
\end{centering}

\caption{\label{cap:bosking} Patterns of orientation columns and long-range
horizontal connections in the primary visual cortex of tree shrew
visualized using optical imaging of intrinsic signals (modified from
\cite{bosking:97}). \textbf{a}, Activity patterns resulting from
stimulation with vertically and obliquely oriented gratings, respectively.
White bars depict the orientation of the visual stimulus. Activated
columns are labeled dark gray. The used stimuli activate only columns
in the primary visual area V1. The patterns thus end at the boundary
between areas V1 and V2. \textbf{b}, The pattern of orientation preferences
calculated from such activity patterns. The orientation preferences
of the columns are color coded as indicated by the bars. A part of
the pattern of orientation preferences is shown at higher magnification.
Two pinwheel centers of opposite topological charge are marked by
arrows. \textbf{c}, Long-range horizontal connections extend over
several millimeters parallel to the cortical surface (tree shrew,
superimposed on the orientation preference map). White symbols indicate
locations of cells that were filled by a tracer (biocytin); labeled
axons are indicated by black symbols.}
\end{figure}

In the visual cortex, as in most areas of the cerebral cortex, information
is processed in a 2-dimensio\-nal (2D) array of functional modules,
called cortical columns \cite{levay/nelson:91,creutzfeldt:95}. Individual
columns are groups of neurons extending vertically throughout the
entire cortical thickness that share many functional properties. Orientation
columns in the visual cortex are composed of neurons preferentially
responding to visual contours of a particular stimulus orientation
\cite{hubel:62}. In a plane parallel to the cortical surface, neuronal
selectivities vary systematically, so that columns of similar functional
properties form highly organized 2D patterns, known as functional
cortical maps (Fig. \ref{cap:bosking}a, b). 

Experimentally, the pattern of orientation preferences can be visualized
using optical imaging methods \cite{blasdel:86,bonhoeffer:91}. Optical
imaging of intrinsic signals is based on the fact that the optical
properties differ in active vs. less active parts of the cortex \cite{bonhoeffer:96}.
This is utilized to record patterns of activity from light reflectance.
In a typical experiment, the activity patterns $E_{k}(\mathbf{x})$
produced by stimulation with a grating of orientation $\theta_{k}$
are recorded. Here $\mathbf{x}=(x,y)$ represents the location of
a column in the cortex. Using the activity patterns $E_{k}(\mathbf{x})$,
a field of complex numbers $z(\mathbf{x})$ can be constructed that
completely describes the pattern of orientation columns: \begin{equation}
z(\mathbf{x})=\sum_{k}e^{i\,2\,\theta_{k}}\, E_{k}(\mathbf{x})\,.\label{eq:orientation-map-experm}\end{equation}
 The pattern of orientation preferences $\vartheta(\mathbf{x})$ is
then obtained from $z(\mathbf{x})$ as follows: \begin{equation}
\vartheta(\mathbf{x})=\frac{1}{2}\, arg(z)\,.\label{eq:ori-pref-experm}\end{equation}
 Typical examples of such activity patterns $E_{k}(\mathbf{x})$ and
the patterns of orientation preferences derived from them are shown
in Fig. \ref{cap:bosking}a, b. Numerous studies confirmed that the
orientation preference of columns is an almost everywhere continuous
function of their position in the cortex. The domains formed by neighboring
columns with similar orientation preference are called iso-orientation
domains \cite{swindale:85}.

\subsection{Pinwheels}

At many locations the iso-orientation domains are arranged radially
around a common center \cite{bonhoeffer:91,bonhoeffer:93}. Around
these pinwheel \cite{braitenberg:79} centers, stimulus orientations
are represented in circular fashion. Such an arrangement had been
previously hypothesized on the basis of electrophysiological experiments
\cite{albus:75,swindale/:87} and theoretical considerations \cite{Swindale:82}.
The regions exhibiting this kind of radial arrangement were termed
pinwheels \cite{braitenberg:79}. The centers of pinwheels are point
discontinuities of the field $\vartheta(\mathbf{x})$ where the mean
orientation preference of nearby columns changes abruptly. They can
be characterized by a topological charge \begin{equation}
q_{i}=\frac{1}{2\pi}\,\oint_{C_{j}}\mathbf{\nabla}\vartheta(\mathbf{x})d\mathbf{s}\label{eq:pinwheel-charge}\end{equation}
which indicates in particular whether the orientation preference increases
clockwise around the center of the pinwheel or counterclockwise. Here,
$C_{j}$ is a closed curve around a single pinwheel center at $\mathbf{x}_{i}$.
Since $\vartheta$ is a cyclic variable within the interval $[0,\pi)$
and up to isolated points is a smooth function of $\mathbf{x}$, $q_{i}$
can only have the values \begin{equation}
q_{i}=\frac{n}{2}\label{eq:charge-+-}\end{equation}
 where $n$ is an integer number \cite{mermin:79}. If its absolute
value $|q_{i}|$ is $1/2$, each orientation is represented once in
the vicinity of a pinwheel center. Pinwheel centers with a topological
charge of $\pm1/2$ are simple zeros of $z(\mathbf{x})$. In experiments
only pinwheels that had the lowest possible topological charge $q_{i}=\pm1/2$
are observed. This means there are only two types of pinwheels: those
whose orientation preference increases clockwise and those whose orientation
preference increases counterclockwise. This organization has been
confirmed in a large number of species and is therefore believed to
be a general feature of visual cortical orientation maps \cite{bonhoeffer/etal:95,bartfeld/grinvald:92,blasdel:92b,blasdel/etal:93,bosking:97,xu:05}.

\subsection{Hypercolumn and pinwheel density}

The pattern of preferred orientations is roughly repetitive \cite{hubel:62,levay/nelson:91}.
The column spacing $\Lambda$, i.e. the spacing between adjacent iso-orientation
domains preferring the same stimulus orientation, is typically in
the range of $\sim1$mm. The column spacing $\Lambda$ determines
the size of the cortical hypercolumn, which is considered to be the
basic processing unit of the visual cortex \cite{hubel:62,creutzfeldt:95,kaschube:02}.
We define the size of a hyper column by $\Lambda^{2}$. The pinwheel
density is defined as the number of pinwheels per unit area $\Lambda^{2}$.
Thus, by this definition, the pinwheel density is independent of the
spacing of columns and dimension-less.

\subsection{Intra-cortical connectivity}

Visual cortical neurons are embedded in densely connected networks
\cite{braitenberg/schutz:98}. Besides a strong connectivity vertical
to the cortical sheet between neurons from different layers within
a column, neurons also form extensive connections horizontal to the
cortical surface linking different orientation columns. These connections
extend for several millimeters parallel to the cortical surface and
are therefore called long-range horizontal connections. As shown in
Fig. \ref{cap:bosking}b for the example of the tree shrew, these
connections are clustered primarily connecting domains of similar
orientation preference. They have been observed in various mammals
\cite{loewel/singer:90,bosking:97,sincich:01,white:01} and repeatedly
hypothesized to play an important to role in visual processing tasks
such as contour integration.

\subsection{Activity-dependent development}

In normal development, orientation columns first form at about the
time of eye opening \cite{chapman:96,crair:98,white:01} which for
the ferret is approximately at post natal day (PD) 31. This is just
a few days after neurons first respond to visual stimuli. A subset
of neurons show orientation preference from that time on, but the
adult pattern is not attained until seven weeks after birth \cite{chapman/stryker:93}.
Roughly clustered horizontal connections are present by around PD
27 \cite{ruthazer/stryker:96}. Many lines of evidence suggest that
the formation of orientation columns is a dynamical process dependent
on neuronal activity and sensitive to visual experience \cite{swindale:96,sur:01}.
This is suggested not only by the time line of normal development,
but also receives support from various experiments manipulating the
sensory input to the cortex. Most intriguingly, when visual inputs
are rewired to drive what would normally become primary auditory cortex,
orientation selective neurons and a pattern of orientation columns
even forms in this brain region that would normally not at all be
involved in the processing of visual information \cite{sharma:00,sur:01}.
This observation suggests that the capability to form a system of
orientation columns is intrinsic to the learning dynamics of the cerebral
cortex given appropriate inputs. Moreover, the comparison of development
under conditions of modified visual experience demonstrates that adequate
visual experience is essential for the complete maturation of orientation
columns and that impaired visual experience, as with experimentally
closed eye-lids \cite{crair:98,white:01} or by rearing kittens in
a striped environment consisting of a single orientation \cite{sengpiel:99},
can suppress or impair the formation of orientation columns (but see
also \cite{goedecke:96}). 

Viewed from a dynamical systems perspective, the activity-dependent
remodeling of the cortical network is a process of dynamical pattern
formation. In this picture, spontaneous symmetry breaking in the developmental
dynamics of the cortical network underlies the emergence of cortical
selectivities such as orientation preference \cite{miller:94}. The
subsequent convergence of the cortical circuitry towards a mature
pattern of selectivities can be viewed as the development towards
an attractor of the developmental dynamics \cite{wolf:98}. This is
consistent with the interpretation of cortical development as an optimization
process. In the following, we will briefly describe a model \cite{wolf:05}
that is based on this view.

\subsection{Modeling cortical self-organization }

Self-organization has been observed to robustly produce large scale
structures in various complex systems. Often, the class of patterns
emerging depends on fundamental system properties such as symmetries
rather than on system specific details. Pattern formation can therefore
often be described by abstract models incorporating these properties
only. For systems undergoing a Turing type instability, canonical
model equations are of the Swift-Hohenberg\cite{swift:77,cross:93}
type \begin{eqnarray}
\partial_{t}z & = & F[z]\nonumber \\
 & = & L_{SH}z+N_{2}[z]+N_{3}[z]+\cdots\label{eq:model-general}\end{eqnarray}
 where the linear part is\begin{equation}
L_{SH}=r-\left(k_{c}^{2}+\nabla^{2}\right)^{2}\label{eq:sh-operator-general}\end{equation}
and $z(\mathbf{x},t)$ is a complex scalar field. If the bifurcation
parameter $r<0$, the homogeneous state $z(\mathbf{x})=0$ is stable.
For $r>0$, a pattern with wavelength close to $\Lambda=2\pi/k_{c}$
emerges. The lowest order nonlinearities $N_{2}$ and $N_{3}$ are
quadratic and cubic in $z$, respectively. The form of these nonlinearities
determines the class of the emerging pattern, i.e. whether hexagons,
rotating stripes, spiral waves, or another type of pattern emerges. 

Following this paradigm, we adopted a model of the form Eq. (\ref{eq:model-general})
with nonlinearities derived from key features of the visual cortex
(following \cite{wolf:05,wolf:05b}). As in experimental recordings,
orientation columns are represented by a complex field \cite{Swindale:82,wolf:98}
\begin{equation}
z(\mathbf{x})=|z(\mathbf{x})|\, e^{i2\vartheta(\mathbf{x})}\,\label{eq:field}\end{equation}
where $\vartheta$ is the orientation preference and $|z|$ a measure
of selectivity at location $\mathbf{x}=(x,y)$ in the map. The factor
$2$ in the exponent accounts for the $\pi$-periodicity of stimulus
parameter orientation. Constructing the nonlinearity of the model
relies on the following assumptions. The model includes the effects
of long-range intracortical connections between columns with similar
orientation preference (Fig. \ref{cap:bosking}c). Unlike many non-living
systems in which interactions are purely local, long-range interactions
are an important and distinctive feature of neuronal circuits in the
cortex and particularly in the primary visual cortex. Further, based
on the spatial homogeneity of circuits across cortex, it is assumed
that the dynamics is symmetric with respect to translations,\begin{equation}
F[\hat{T}_{\mathbf{y}}\, z]=\hat{T}_{\mathbf{y}}\, F[z]\;\text{ with }\:\hat{T}_{\mathbf{y}}\, z(\mathbf{x})=z(\mathbf{x}+\mathbf{y})\,,\label{eq:transl-sym}\end{equation}
and rotations\begin{equation}
F[\hat{R}_{\beta}\, z]=\hat{R}_{\beta}\, F[z]\;\text{ with }\;\hat{R}_{\beta}\, z(\mathbf{x})=z\left({\left[\begin{array}{ll}
\cos(\beta) & \sin(\beta)\\
-\sin(\beta) & \cos(\beta)\end{array}\right]}\,\mathbf{x}\right)\label{eq:rotation-sym}\end{equation}
 of the cortical sheet. This means that patterns that can be converted
into one another by translation or rotation of the cortical layers
belong to equivalent solutions of the model, Eq. (\ref{eq:model-general})
, by construction. It is further assumed that the dynamics is symmetric
with respect to shifts in orientation\begin{equation}
F[e^{i\phi}\, z]=e^{i\phi}\, F[z]\,.\label{eq:shift-symm}\end{equation}
 Thus, two patterns are also equivalent solutions of the model, if
their layout of orientation domains is identical, but the preferred
orientations differ everywhere by the same constant angle. Solutions
shall contain representations of all stimulus orientations. For simplicity,
couplings to other visual cortical representations such as ocular
dominance or retinotopy are neglected. Considering only leading order
terms up to cubic nonlinearities a nonlinearity fulfilling these requirements
is given by \begin{eqnarray}
N_{2}[z(\mathbf{x})] & = & 0\nonumber \\
N_{3}[z(\mathbf{x})] & = & \left(1-g\right)|z(\mathbf{x})|^{2}z(\mathbf{x})+\nonumber \\
 &  & \left(g-2\right)\int d^{2}\mathbf{y}\, K_{\sigma}\left(\mathbf{y}-\mathbf{x}\right)\left(z(\mathbf{x})|z(\mathbf{y})|^{2}+\frac{1}{2}\bar{z}(\mathbf{x})z(\mathbf{y})^{2}\right)\,,\label{eq:nonlinp-model-general}\end{eqnarray}
with long-range interactions mediated through convolutions with a
Gaussian \begin{equation}
K_{\sigma}(\mathbf{x})=\frac{1}{2\pi\sigma^{2}}e^{-\frac{\mathbf{x}^{2}}{2\sigma^{2}}}\label{eq:Gauss-lrc}\end{equation}
with range $\sigma$. The second parameter $0\le g\le2$ controls
local and nonlocal interactions. The first term is the only strictly
local term consistent with the required symmetries, the second non-local
term represents the simplest non-local term that is symmetric with
respect to these and with respect to permutations

\begin{equation}
N_{3}(u,v,w)=N_{3}(w,u,v)\,.\label{eq:permutation-sym}\end{equation}
Here the non-linear operator is written in a trilinear form as introduced
in \cite{wolf:05,wolf:05b}. This additional symmetry implies that
all two-orientation solutions, for instance real valued solutions,
of the model, are unstable, which in turn guarantees that all stimulus
orientations are represented. One should note that in this model the
spatial range of the nonlinearity $\sigma$ is a control parameter
independent of the wavelength $\Lambda$. The patterns selected for
different ratios $\sigma/\Lambda$ are displayed in Fig. \ref{cap:ess-compl-planf}b.
It can also be derived as an approximation to a model for the combined
development of orientation preference and long-range horizontal connections
\cite{wolf:05b}. The model as defined in Eq. (\ref{eq:model-general}-\ref{eq:permutation-sym})
is variational. It is consistent with synaptic models based on Hebbian
plasticity, e.g. \cite{swindale:96,erwin:95,miller:99}. It is the
only model known to the authors that exhibits stable aperiodic solutions
dominated by a single spatial scale $\Lambda$. \textbf{}These solutions
resemble orientation maps observed in the visual cortex. In the following,
we discuss the structure of these solutions and the phase diagram
of the model.

\subsection{Weakly nonlinear analysis}

Solutions of the model close to the bifurcation point $r=0$ are known
in closed form, derived by means of a perturbation method called weakly
nonlinear stability analysis\cite{manneville:1990,cross:93}. When
the dynamics is close to a finite wavelength instability, the essential
Fourier components of the emerging pattern are located on the critical
circle. Solutions are planform patterns \begin{equation}
z(\mathbf{x})=\sum_{j}A_{j}e^{i\mathbf{k}_{j}\mathbf{x}}\label{eq:planform-general}\end{equation}
 composed of a finite number of Fourier components with wavevectors
on the critical circle, $|\mathbf{k}_{j}|=k_{c}$. By symmetry, the
dynamics of amplitudes $A_{i}$ of a planform are governed by amplitude
equations\begin{equation}
\dot{A}_{i}=A_{i}-\sum_{j}g_{ij}\left|A_{j}\right|^{2}A_{i}-\sum_{j}f_{ij}A_{j}A_{j^{-}}\bar{A}_{i^{-}}\label{eq:amp-eqn-general}\end{equation}
where $j^{-}$ denotes the index of the mode antiparallel to mode
$j$. The form of Eq. (\ref{eq:amp-eqn-general}) is universal for
models of a complex field $z$ satisfying symmetry assumptions (\ref{eq:transl-sym}-\ref{eq:shift-symm}).
All model dependencies are included in the coupling coefficients $g_{ij}$
and $f_{ij}$ and may be obtained from $F[z]$ by multiscale expansion
\cite{manneville:1990,cross:93}. Denoting the angle between the wave
vectors $\mathbf{k}_{i}$ and $\mathbf{k}_{j}$ by $\alpha$ and $\delta_{ij}$
the Kronecker delta\textbf{,} the coefficients read\cite{wolf:05,wolf:05b}
\begin{eqnarray}
g_{ij} & = & \left(1-\frac{1}{2}\delta_{ij}\right)g\left(\alpha\right)\nonumber \\
f_{ij} & = & \left(1-\delta_{ij}-\delta_{i^{-}j}\right)f\left(\alpha\right)\label{eq:interaction-coeff}\end{eqnarray}
where\begin{eqnarray}
g(\alpha) & = & g+2(2-g)\exp\left(-\sigma^{2}k_{c}^{2}\right)\cosh\left(\sigma^{2}k_{c}^{2}\cos(\alpha)\right)\nonumber \\
f(\alpha) & = & \frac{1}{2}g(\alpha)\label{eq:interaction-function}\end{eqnarray}
are called angle-dependent interaction functions. 

\begin{figure}
\begin{centering}\includegraphics[width=1\textwidth]{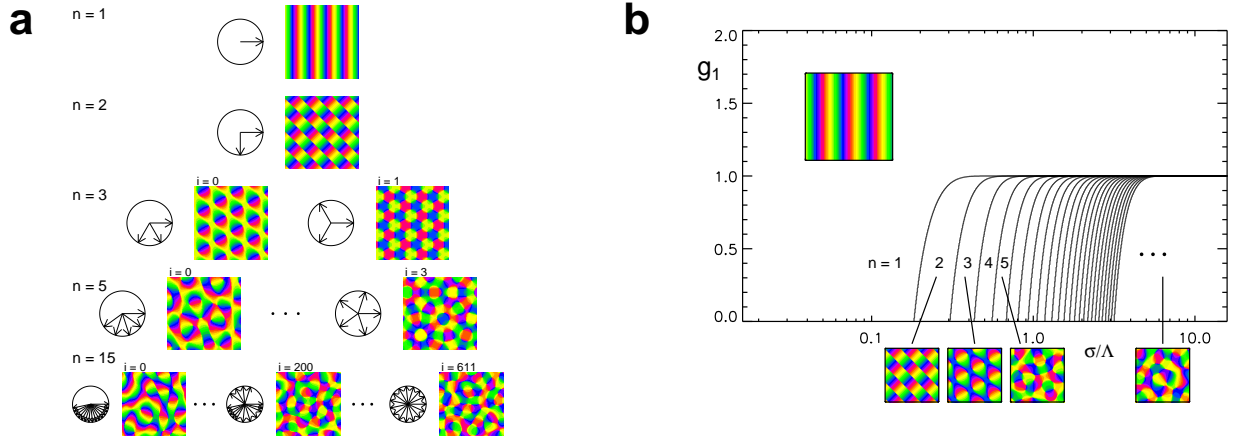}\par\end{centering}

\caption{\label{cap:ess-compl-planf} \textbf{a}, Essentially complex planforms
with different numbers $n=1,2,3,5,15$ of active modes: The patterns
of orientation preferences $\theta(\mathbf{x})$ are shown. The diagrams
to the left of each pattern display the position of the wavevectors
of active modes on the critical circle. For $n=3$, there are two
patterns; for $n=5$, there are four; and for $n=15$, there are $612$
different patterns. \textbf{b}, Phase diagram of model. If non-local
interactions are dominant ($g<1$) and long-ranging ($\sigma$ large
compared to $\Lambda$), quasiperiodic planforms are selected. Reproduced
from \cite{wolf:05b}.}
\end{figure}

Stationary solutions of Eq. (\ref{eq:amp-eqn-general}) are given
by families of planforms\cite{wolf:05,wolf:05b} 

\begin{equation}
z(\mathbf{x})=\sum_{j=0}^{n-1}\left|A_{j}\right|e^{i(l_{j}\mathbf{k}_{j}\mathbf{x}+\phi_{j})}\label{eq:planforms-specific}\end{equation}
 of order $n$ with wavevectors \begin{equation}
\mathbf{k}_{j}=k_{c}\left(\cos\left(\frac{j\pi}{n}\right),\sin\left(\frac{j\pi}{n}\right)\right)\label{eq:wavevec-definiton-funds}\end{equation}
 distributed equidistantly on the upper half of the critical circle
and binary values $l_{j}=\pm1$ determining whether the mode with
wave vector $\mathbf{k}_{j}$ or with wavevector $-\mathbf{k}_{j}$
is active. These planforms cannot realize a real valued function and
are called essentially complex planforms (Fig. \ref{cap:ess-compl-planf}).
For these planforms the third term in Eq. (\ref{eq:amp-eqn-general})
vanishes and the effective amplitude equations for the active modes
reduce to a system of Landau equations \begin{equation}
\dot{A}_{i}=A_{i}-\sum_{j}g_{ij}\left|A_{j}\right|^{2}A_{i}\label{eq:landou-equations}\end{equation}
with stationary solutions Eq. (\ref{eq:planforms-specific}) with
amplitudes of equal modulus\begin{equation}
\left|A_{i}\right|=\left(\sum_{j}g_{ij}\right)^{-1/2}\label{eq:amplitude-ess-compl-plnf}\end{equation}
and an arbitrary phase $\phi_{i}$ independent of the mode configuration
$l_{j}$. If the dynamics is stabilized by long-range nonlocal interactions
($g<1$, $\sigma>\Lambda$), large $n$ planforms are the only stable
solutions. In this long-range regime, the order of planforms grows
as \begin{equation}
n\sim2\pi\sigma/\Lambda\,,\label{eq:sigma_lambda}\end{equation}
approximately linear with the interaction range. For a given order
$n$, different planforms are degenerated in energy. This is a consequence
of the permutation symmetry Eq. (\ref{eq:permutation-sym}). This
symmetry also implies that the relevant stable solutions are essentially
complex planforms which in turn guarantees that all stimulus orientations
are represented.

\section{Calculation of pinwheel density \label{sec:amplitude}}

\subsection{Large range limit of interactions: Planform anisotropy}

First, we calculate the average pinwheel density $\rho_{l}$ for an
ensemble of planforms, Eq. (\ref{eq:planforms-specific}), with a
fixed set of wavevector directions $l=\left(l_{0},l_{1},\dots,l_{n-1}\right)$
but arbitrary phases $\phi_{j}$ in the limit $n\rightarrow\infty$.
Here, and until noted otherwise, $\left\langle \right\rangle $ shall
denote average over phases $\phi_{j}$. In this limit, $z$ and local
linear functionals of $z$ have Gaussian statistics such that the
density of pinwheels is determined by the second order statistics
of the field. Second, we evaluate the expectation value of $\rho_{l}$
over all sets of $l$. 

For large $\sigma$, to good approximation $g_{ii}\approx1$ and $g_{ij}\approx g$
and hence $\left|A_{i}\right|\approx1/\sqrt{ng}$. Planforms (\ref{eq:planforms-specific})
simplify to \begin{equation}
z(\mathbf{x})=\sqrt{\frac{2}{n}}\sum_{j=0}^{n-1}e^{i(l_{j}\mathbf{k}_{j}\mathbf{x}+\phi_{j})}\label{eq:planform-solution}\end{equation}
where for later convenience the constant $\sqrt{2g}$ was absorbed
into $z(\mathbf{x})$. Pinwheels are the zeros of the field $z(\mathbf{x})$.
The number of pinwheels in a given area $A$ is obtained by \begin{equation}
N=\int_{A}d^{2}\mathbf{x}\,\delta(z(\mathbf{x}))\, J(z(\mathbf{x}))\,,\label{eq:pinwheel-number}\end{equation}
 where $\delta(\mbox{x})$ denotes Dirac's delta function and \begin{equation}
J(z(\mathbf{x}))=\left|\frac{\partial R(\mathbf{x})}{\partial x}\frac{\partial I(\mathbf{x})}{\partial y}-\frac{\partial R(\mathbf{x})}{\partial y}\frac{\partial I(\mathbf{x})}{\partial x}\right|\label{eq:jacob-determ}\end{equation}
is the Jacobian of the field \begin{equation}
z(\mathbf{x})=R(\mathbf{x})+iI(\mathbf{x})\label{eq:split-real-imag}\end{equation}
split here for later convenience into its real and imaginary part.
Averaging Eq. (\ref{eq:pinwheel-number}) over the ensemble of phases
$\phi_{j}$ reads\begin{equation}
\langle N\rangle=\int_{A}d^{2}\mathbf{x}\left\langle \delta(z(\mathbf{x}))J(z(\mathbf{x}))\right\rangle \label{eq:pinwheel-numb-ensemble}\end{equation}
implying that 

\begin{equation}
\rho_{l}=\left\langle \delta(z(\mathbf{x}))J(z(\mathbf{x}))\right\rangle \label{eq:pinwheel-dens-def}\end{equation}
is the expectation value of the pinwheel density for a fixed set of
$l$. This expectation value only depends on local quantities, namely
on the field, Eq. (\ref{eq:planform-solution}), and its spatial derivatives
\begin{equation}
\nabla z(\mathbf{x})=i\sqrt{\frac{2}{n}}\sum_{j=0}^{n-1}l_{j}\mathbf{k_{j}}e^{i(l_{j}\mathbf{k}_{j}\mathbf{x}+\phi_{j})}\label{eq:planform-solution-deriv}\end{equation}
such that knowledge of the joint probability density $p\left(z,\nabla z\right)$
is sufficient to evaluate Eq. (\ref{eq:pinwheel-dens-def}). Owing
to the central limit theorem, this probability density becomes Gaussian
in the large $n$ limit. Eq. (\ref{eq:pinwheel-dens-def}) is then
determined by the first and second order statistics of $z$ and $\nabla z$.
Furthermore, since their statistics is the same at each location $\mathbf{x}$,
it is sufficient to evaluate Eq. (\ref{eq:pinwheel-dens-def}) for
$z(\mathbf{0})$, $\nabla z(\mathbf{0})$. The spatial dependency
is thus omitted in the following. The average in Eq. (\ref{eq:pinwheel-dens-def})
is given by an integral over the joint probability density\begin{equation}
p(v)=\frac{1}{(2\pi)^{3}\sqrt{\det C}}\, e^{-\frac{1}{2}v^{T}C^{-1}v}\label{eq:jpd}\end{equation}
of components \begin{equation}
\mathbf{v}=\left(R,I,\partial_{x}R,\partial_{x}I,\partial_{y}R,\partial_{y}I\right)\,.\label{eq:v-vector}\end{equation}
 with covariance matrix $C$ which shall be analyzed in the following. 

First, the diagonal elements of $C$ are evaluated. Using (\ref{eq:planforms-specific})
and (\ref{eq:split-real-imag}), the auto-correlations of the field
are \begin{eqnarray}
\left\langle R^{2}\right\rangle  & = & \frac{2}{n}\sum_{j,j'=0}^{n-1}\left\langle \cos\phi_{j}\cos\phi_{j'}\right\rangle \nonumber \\
 & = & \textrm{1}\label{eq:rr-corr}\end{eqnarray}
 for the real part and\begin{eqnarray}
\left\langle I^{2}\right\rangle  & = & \frac{2}{n}\sum_{j,j'=0}^{n-1}\left\langle \sin\phi_{j}\sin\phi_{j'}\right\rangle \nonumber \\
 & = & \textrm{1}\label{eq:ii-corr}\end{eqnarray}
for the imaginary part. For the spatial derivatives one obtains\begin{eqnarray}
\left\langle \left(\partial_{x}R\right)^{2}\right\rangle  & = & \frac{2}{n}\sum_{j,j'=0}^{n-1}l_{j}l_{j'}k_{xj}k_{xj'}\left\langle \sin\phi_{j}\sin\phi_{j'}\right\rangle \nonumber \\
 & = & \frac{1}{n}\sum_{j=0}^{n-1}k_{xj}^{2}\label{eq:rxrx-corr-def}\end{eqnarray}
where $k_{xj}$ is the $x$-component of $\mathbf{k}_{j}$ and, likewise,\begin{align}
\left\langle \left(\partial_{x}I\right)^{2}\right\rangle  & =\frac{1}{n}\sum_{j}k_{xj}^{2}\nonumber \\
\left\langle \left(\partial_{y}R\right)^{2}\right\rangle  & =\frac{1}{n}\sum_{j}k_{yj}^{2}\nonumber \\
\left\langle \left(\partial_{y}I\right)^{2}\right\rangle  & =\frac{1}{n}\sum_{j}k_{yj}^{2}\,.\label{eq:rxrx-rest}\end{align}
The equality of these correlations follows from inserting Eq. (\ref{eq:wavevec-definiton-funds})
into (\ref{eq:rxrx-corr-def}) yielding \begin{eqnarray}
\left\langle \left(\partial_{x}R\right)^{2}\right\rangle  & = & \frac{k_{c}^{2}}{n}\sum_{j=0}^{n-1}\cos^{2}\left(\frac{j\pi}{n}\right)\nonumber \\
 & = & \frac{k_{c}^{2}}{n}\sum_{j=0}^{n-1}\sin^{2}\left(\frac{j\pi}{n}-\frac{\pi}{2}\right)\nonumber \\
 & = & \frac{k_{c}^{2}}{n}\sum_{j=0}^{n-1}\sin^{2}\left(\frac{j\pi}{n}\right)\nonumber \\
 & = & \left\langle \left(\partial_{y}R\right)^{2}\right\rangle \label{eq:calc1}\end{eqnarray}
and also\begin{eqnarray}
\left\langle \left(\partial_{x}R\right)^{2}\right\rangle  & = & \frac{k_{c}^{2}}{n}\sum_{j=0}^{n-1}\cos^{2}\left(\frac{j\pi}{n}\right)\nonumber \\
 & = & k_{c}^{2}-\frac{k_{c}^{2}}{n}\sum_{j=0}^{n-1}\sin^{2}\left(\frac{j\pi}{n}\right)\nonumber \\
 & = & k_{c}^{2}-\left\langle \left(\partial_{x}R\right)^{2}\right\rangle \label{eq:calc2}\end{eqnarray}
such that all auto-correlations become \begin{equation}
\left\langle \left(\partial_{x}R\right)^{2}\right\rangle =\left\langle \left(\partial_{y}R\right)^{2}\right\rangle =\left\langle \left(\partial_{x}I\right)^{2}\right\rangle =\left\langle \left(\partial_{y}I\right)^{2}\right\rangle =\frac{k_{c}^{2}}{2}=2\pi^{2}\label{eq:rxrx-all-corr}\end{equation}
when choosing without loss of generality the column spacing to be
$\Lambda=2\pi/k_{c}\equiv1$.

Most off-diagonal elements of the covariance matrix $C$ vanish. All
non-vanishing contributions are related to the planform anisotropy
defined by\begin{equation}
\vec{\xi}\equiv\frac{1}{4n}\sum_{j=0}^{n-1}l_{j}\mathbf{k}_{j}\label{eq:aniso-def}\end{equation}
 which depends on the set $l$. The covariance between the field and
its derivative reads\begin{eqnarray}
\left\langle z\nabla\bar{z}\right\rangle  & = & -\frac{2}{n}i\sum_{j,j'=0}^{n-1}l_{j'}\mathbf{k}_{j'}\left\langle e^{i(\phi_{j}-\phi_{j'})}\right\rangle \nonumber \\
 & = & -\frac{2}{n}i\sum_{j=0}^{n-1}l_{j}\mathbf{k}_{j}\nonumber \\
 & = & -2i\vec{\chi}\label{eq:corr-z-grad}\end{eqnarray}
with $\vec{\chi}\equiv\frac{1}{n}\sum_{j=0}^{n-1}l_{j}\mathbf{k}_{j}\ge0$.
The modulus $\chi=\left|\vec{\chi}\right|$ is small for an isotropic
distribution of wavevectors $l_{j}\mathbf{k}_{j}$. To estimate its
upper bound $\chi_{max}$, consider the most anisotropic case with
all $l_{j}\mathbf{k}_{j}$ situated in the right plane ($l_{j}=1$
for $j\le n/2$, $l_{j}=-1$ for $j>n/2$). For large $n$ this upper
bound is 

\begin{eqnarray}
\chi_{max} & = & \left|\frac{1}{n}\sum_{j=0}^{n-1}l_{j}\mathbf{k}_{j}\right|\nonumber \\
 & = & \frac{k_{c}}{\pi}\left|\sum_{j=-\frac{n}{2}}^{\frac{n}{2}-1}\frac{\pi}{n}e^{i\pi\frac{j}{n}}\right|\nonumber \\
 & \approx & \frac{k_{c}}{\pi}\left|\int_{-\frac{\pi}{2}}^{\frac{\pi}{2}}d\alpha e^{i\alpha}\right|\nonumber \\
 & = & 4\label{eq:calc3}\end{eqnarray}
such that the modulus $\xi=\left|\vec{\xi}\right|$ of the anisotropy
is bounded by $0\le\xi\le1$. In the following, without loss of generality,
we assume $\vec{\xi}=\xi(1,0)$ implying that all correlations involving
one derivative in $y$-direction vanish. Correlations involving $\partial_{x}$
are obtained by writing Eq. (\ref{eq:corr-z-grad}) and

\begin{eqnarray}
\left\langle z\nabla z\right\rangle  & = & \frac{2}{n}i\sum_{j,j'=0}^{n-1}l_{j'}k_{j'}\left\langle e^{i(\phi_{j}+\phi_{j'})}\right\rangle \nonumber \\
 & = & 0\label{eq:ccorr-z-grad}\end{eqnarray}
in the form\begin{alignat}{1}
\left\langle z\nabla_{x}\bar{z}\right\rangle  & =\left\langle R\partial_{x}R\right\rangle +\left\langle I\partial_{x}I\right\rangle +i\left(\left\langle I\partial_{x}R\right\rangle -\left\langle R\partial_{x}I\right\rangle \right)=-i8\xi\nonumber \\
\left\langle z\nabla_{x}z\right\rangle  & =\left\langle R\partial_{x}R\right\rangle -\left\langle I\partial_{x}I\right\rangle +i\left(\left\langle I\partial_{x}R\right\rangle +\left\langle R\partial_{x}I\right\rangle \right)=0\label{eq:both-corr-zgradz}\end{alignat}
and comparing both imaginary parts showing that \begin{equation}
-\left\langle I\partial_{x}R\right\rangle =\left\langle R\partial_{x}I\right\rangle =4\xi\label{eq:corr-irx-rix}\end{equation}
does not vanish for anisotropic planforms. 

Expression (\ref{eq:corr-irx-rix}) are the only non-vanishing non-diagonal
elements of the matrix $C$. Indeed, \begin{equation}
\left\langle R\partial_{x}R\right\rangle =\left\langle I\partial_{x}I\right\rangle =0\,,\label{eq:corr-rrx-iix}\end{equation}
follows from comparing both real parts in Eq. (\ref{eq:both-corr-zgradz}).
Furthermore, correlations between the real and imaginary part and
between their derivatives, e.g. \begin{alignat}{1}
\left\langle RI\right\rangle  & =0\nonumber \\
\left\langle \partial_{x}R\partial_{x}I\right\rangle  & =0\nonumber \\
\left\langle \partial_{y}R\partial_{x}I\right\rangle  & =0\,,\label{eq:cross-real-imag}\end{alignat}
 vanish since they contain terms of the form$\left\langle \sin\phi_{j}\cos\phi_{j'}\right\rangle =\left\langle \sin\phi_{j}\right\rangle \left\langle \cos\phi_{j}\right\rangle =0$.
Finally, because \begin{eqnarray}
\left\langle \partial_{x}R\partial_{y}R\right\rangle  & = & \frac{2}{n}\sum_{jj'}l_{j}l_{j'}k_{jx}k_{j'y}\left\langle \sin\phi_{j}\sin\phi_{j'}\right\rangle \nonumber \\
 & = & \frac{1}{n}\sum_{j}k_{jx}k_{jy}\nonumber \\
 & = & \frac{2k_{c}^{2}}{n}\sum_{j=0}^{n-1}\sin\left(2\pi\frac{j}{n}\right)\label{eq:calc4}\end{eqnarray}
vanishes for arbitrary $n$, also correlations between derivatives
in different directions, \begin{equation}
\left\langle \partial_{x}R\partial_{y}R\right\rangle =\left\langle \partial_{x}I\partial_{y}I\right\rangle =0\,,\label{eq:cross-rx-ry}\end{equation}
do not contribute to the density of pinwheels. 

Altogether, the covariance matrix for the vector $\mathbf{v}=(R,I,\partial_{x}R,\partial_{x}I,\partial_{y}R,\partial_{y}I)$
reads 

\begin{eqnarray}
C & = & \left(\begin{array}{cccccc}
\langle R^{2}\rangle &  &  & \langle\partial_{x}IR\rangle\\
 & \langle I^{2}\rangle & \langle\partial_{x}RI\rangle\\
 & \langle I\partial_{x}R\rangle & \langle(\partial_{x}R)^{2}\rangle\\
\langle R\partial_{x}I\rangle &  &  & \langle(\partial_{x}I)^{2}\rangle\\
 &  &  &  & \langle(\partial_{y}R)^{2}\rangle\\
 &  &  &  &  & \langle(\partial_{y}I)^{2}\rangle\end{array}\right)\nonumber \\
 & = & \left(\begin{array}{cccccc}
1 &  &  & 4\xi_{l}\\
 & 1 & -4\xi_{l}\\
 & -4\xi_{l} & 2\pi^{2}\\
4\xi_{l} &  &  & 2\pi^{2}\\
 &  &  &  & 2\pi^{2}\\
 &  &  &  &  & 2\pi^{2}\end{array}\right)\,.\label{eq:matrix}\end{eqnarray}

The integral we want to solve is \begin{equation}
\rho_{l}=\frac{1}{(2\pi)^{3}\sqrt{\det\, C}}\int d^{6}\mathbf{v}\,\delta(R)\delta(I)\, J\, e^{-\frac{1}{2}\mathbf{v}^{T}C^{-1}\mathbf{v}}\,.\label{eq:integral}\end{equation}
For the exponent \begin{equation}
E=-\frac{1}{2}\mathbf{v}^{T}C^{-1}\mathbf{v}\label{eq:exponent}\end{equation}
we find

\begin{equation}
E=-\frac{1}{4}\left(\frac{{(\partial_{y}I)}^{2}+(\partial_{y}R)^{2}}{\pi^{2}}+\frac{(\partial_{x}I)^{2}+(\partial_{x}R)^{2}+2\pi^{2}(R^{2}+I^{2})+8\xi\,(I\,\partial_{x}R-R\,\partial_{x}I)}{\pi^{2}-8\xi^{2}}\right)\,.\label{eq:exponent_eval}\end{equation}
Performing the integral over $R$ and $I$, we get

\begin{equation}
\rho_{l}=\frac{1}{(2\pi)^{3}\sqrt{\det\, C}}\int d^{4}\mathbf{w}\, J\,\exp-\frac{1}{4}\left(\frac{(\partial_{x}R)^{2}+(\partial_{x}I)^{2}}{\pi^{2}-8\xi^{2}}+\frac{(\partial_{y}R)^{2}+(\partial_{y}I)^{2}}{\pi^{2}}\right)\label{eq:integral_eval1}\end{equation}
where $\mathbf{w}=(\partial_{x}R,\partial_{x}I,\partial_{y}R,\partial_{y}I)$.
Substituting

\begin{eqnarray}
\partial_{x}R & = & r_{2}\cos\theta_{2}\nonumber \\
\partial_{y}R & = & r_{1}\cos\theta_{1}\nonumber \\
\partial_{x}I & = & r_{2}\sin\theta_{2}\nonumber \\
\partial_{y}I & = & r_{1}\sin\theta_{1}\label{eq:subst}\end{eqnarray}
where \begin{eqnarray}
0 & \le & \theta_{1},\theta_{2}\le2\pi\label{eq:constraints}\\
0 & \le & r_{1},r_{2}<\infty\nonumber \end{eqnarray}
we have \begin{equation}
d^{4}\mathbf{w}=r_{1}r_{2}dr_{1}dr_{2}d\theta_{1}d\theta_{2}\label{eq:vol_elm2}\end{equation}
and the Jacobian reads\begin{equation}
J=r_{1}r_{2}|\sin(\theta_{1}-\theta_{2})|\label{eq:jacobian2}\end{equation}
such that we obtain\begin{equation}
\rho_{l}=\frac{4}{(2\pi)^{2}\sqrt{\det\, C}}\int dr_{1}dr_{2}\, r_{1}^{2}r_{2}^{2}\,\,\exp-\frac{1}{4}(\frac{r_{2}^{2}}{\pi^{2}-8\xi^{2}}+\frac{r_{1}^{2}}{\pi^{2}})\label{eq:integral_eval2}\end{equation}
after integrating over angles. Evaluating the determinant as \begin{equation}
\det\, C=16\pi^{4}(\pi^{2}-8\xi^{2})^{2}\label{eq:det}\end{equation}
and performing the final integration steps, we obtain the pinwheel
density 

\begin{eqnarray}
\rho_{l} & = & \pi\sqrt{1-\frac{8}{\pi^{2}}\xi^{2}}\label{eq:pwd-xi}\end{eqnarray}
of an ensemble of planforms with a fixed set of wavevector directions
$l$. This result shows that the pinwheel density $\rho_{l}$ only
depends on the anisotropy $\xi$ of the considered planform. Since
the anisotropy ranges within $0\le\xi\le1$, the pinwheel density
is confined by $1.36\lesssim\rho_{l}\le\pi$.

\subsection{Distribution of planform anisotropies}

What is the distribution of anisotropies $\xi$ in the large $n$
limit? To address this question consider the ensemble defined by the
different sets $l=\left(l_{0},\dots,l_{n-1}\right)$. From now on,
$\left\langle \right\rangle $ shall denote the expectation value
over this ensemble. In the following we assume that the distribution
of the vector anisotropies $\vec{\xi}$ is isotropic and Gaussian
in the large $n$ limit. The isotropy follows from the rotation symmetry
of the model equations. It implies that \begin{equation}
\left\langle \vec{\xi}\right\rangle =0\,.\label{eq:avg-xi}\end{equation}
The assumption of Gaussian statistics of $ $$\vec{\xi}$ is justified
by the fact that by its definition (\ref{eq:aniso-def}), the vector
anisotropies $\vec{\xi}$ results from various wavevectors $l_{j}\mathbf{k}_{j}$
with pairwise independent directions $l_{j}$. The distribution thus
reads

\begin{equation}
p(\vec{\xi})=\frac{1}{\pi\nu_{\xi}}\exp\left(-\frac{\vec{\xi}^{2}}{\nu_{\xi}}\right)\label{eq:distr-vec-xi}\end{equation}
with variance $\nu_{\xi}$ given by \begin{eqnarray}
\nu_{\xi} & = & \left\langle \vec{\xi}^{2}\right\rangle \nonumber \\
 & = & \frac{1}{16n^{2}}\sum_{jj'}\mathbf{k}_{j}\mathbf{k}_{j'}\left\langle l_{j}l_{j'}\right\rangle \nonumber \\
 & = & \frac{1}{16n^{2}}\sum_{j}\mathbf{k}_{j}^{2}\nonumber \\
 & = & \frac{\pi^{2}}{4n}\,,\label{eq:var-xi}\end{eqnarray}
where $\left\langle l_{j}l_{j'}\right\rangle =\delta_{jj'}$. The
probability density for $\xi$ follows from the distribution of the
vector anisotropy $\vec{\xi}$ by\begin{eqnarray}
p(\xi) & = & 2\pi\xi p\left(\left|\vec{\xi}\right|\right)\nonumber \\
 & = & \frac{8n}{\pi^{2}}\xi\exp\left(-\frac{4n}{\pi^{2}}\xi^{2}\right)\,,\quad\xi\ge0\label{eq:distr-xi}\end{eqnarray}
where in the first equation the prefactor accounts for the change
to polar coordinates.

\subsection{Pinwheel density in the large $n$ limit }

Distribution (\ref{eq:distr-xi}) and Eq. (\ref{eq:pwd-xi}) yield
a distribution of pinwheel density $\rho_{l}$ by means of the coordinate
transform $\xi\rightarrow\rho_{l}$. With

\begin{eqnarray}
p(\rho_{l}) & = & p\left(\xi\left(\rho_{l}\right)\right)\left|\frac{d\xi}{d\rho_{l}}\right|\label{eq:distr-rho-of-xi}\end{eqnarray}
and $\xi=\sqrt{(\pi^{2}-\rho_{l}^{2})/8}$ (Eq. (\ref{eq:pwd-xi}))
the pinwheel density distribution reads \begin{eqnarray}
p(\rho_{l}) & = & \frac{n}{\pi^{2}}\rho_{l}\exp\left(-\frac{n}{2\pi^{2}}\left(\pi^{2}-\rho_{l}^{2}\right)\right)\,,\quad0\le\rho_{l}\le\pi\,.\label{eq:distr-rho}\end{eqnarray}
The expectation value of the pinwheel density follows as 

\begin{eqnarray}
\left\langle \rho\right\rangle  & = & \int_{0}^{\pi}d\rho_{l}\rho_{l}\, p(\rho_{l})\nonumber \\
 & = & \pi-\frac{e^{-\frac{n}{2}}\pi^{\frac{3}{2}}\Phi_{i}\left(\sqrt{\frac{n}{2}}\right)}{\sqrt{2n}}\label{eq:ierf}\end{eqnarray}
where $\Phi_{i}$ is the imaginary error function. In the limit $n\rightarrow\infty$,
the second term vanishes implying 

\begin{equation}
\lim_{n\rightarrow\infty}\left\langle \rho\right\rangle =\pi\:.\label{eq:pwd}\end{equation}
Moreover, the pinwheel density is $\delta$-distributed at $\langle\rho\rangle=\pi$
in the large $n$ limit as its variance $\nu$ converges towards $0$.
Having\begin{eqnarray}
\left\langle \rho^{2}\right\rangle  & = & \int_{0}^{\pi}d\rho\rho^{2}p(\rho)\nonumber \\
 & = & \pi^{2}\frac{2e^{-\frac{n}{2}}+n-2}{n}\label{eq:ierf-2}\end{eqnarray}
which goes to $\pi^{2}$ for $n\rightarrow\infty$, one finds \begin{eqnarray}
\nu & = & \langle\rho^{2}\rangle-\langle\rho\rangle^{2}\nonumber \\
 & = & 0\label{eq:variance}\end{eqnarray}
suggesting that for large $n$ not only the average pinwheel densities
$\langle\rho\rangle$ but the density $\rho$ of almost every realization
is close to $\pi$. Thus, in the limit of infinite interaction range,
the value of the average pinwheel density is $\pi$ in almost every
realization.

\subsection{Intermediate range of interaction }

\begin{figure}
\begin{minipage}[c][1\totalheight][t]{0.5\textwidth}%
\includegraphics[width=0.8\textwidth]{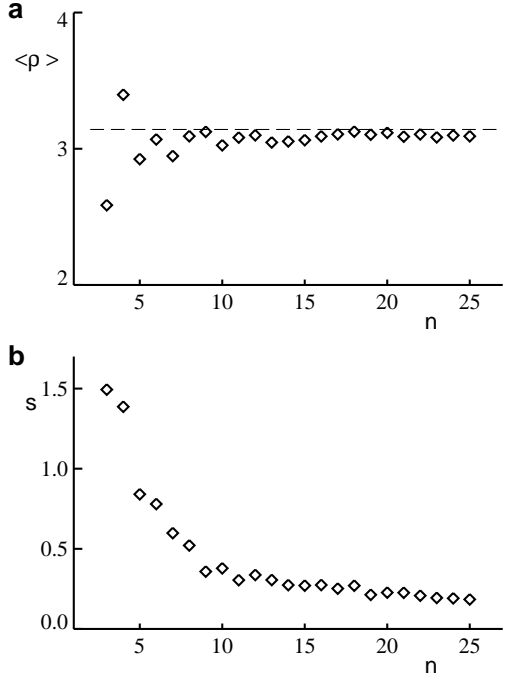}%
\end{minipage}%
\hfill{}%
\begin{minipage}[c][1\totalheight][t]{0.4\textwidth}%

\caption{\label{cap:planform} Pinwheel densities for finite interaction range
$n\approx2\pi\sigma/\Lambda$. \textbf{a}, Average pinwheel density
$\langle\rho\rangle$ in planforms of order $n$ (diamonds). Planforms
with random sets of wavevector directions $l_{j}$ and phases $\phi_{j}$
were synthesized on $2048\times2048$ grids with aspect ratio $\Gamma=32$.
For each planform the pinwheel density was determined in a quadratic
$\Gamma=8$ subregion. Planforms were drawn until a achieving SEM
$\Delta=0.03$ for $n<15$ and $\Delta=0.01$ for $n\geq15$. The
dashed line represents $\langle\rho\rangle=\pi$, valid in the limit
$n\rightarrow\infty$. Note that the average pinwheel density $\langle\rho\rangle$
is close to $\pi$ even at intermediate orders of $n$. \textbf{b},
Standard deviation (SD) $s$ of densities $\rho$. Note that $s$
decreases with order $n$. }%
\end{minipage}%

\end{figure}

Next, we ask to which extent these results do remain valid for finite
interaction ranges and weather they are fairly robust against variations
in range. Real cortices are of course finite and the effective interaction
range for a given species or animal may depend on the range of intracortical
long-range horizontal connections (see Fig. \ref{cap:bosking})c.
It is therefore important to analyze the pinwheel density in solutions
for intermediate interaction range $\sigma$, i.e. in planforms of
intermediate order $n$. To this end, we numerically synthesize planforms
of various order $n$ with randomly chosen $l_{j}$ and $\phi_{j}$
and calculate the pinwheel density averaged across planforms. Pinwheel
centers are identified by the crossings of the zero contour lines
of the real and imaginary part of the field $z$. To calculate the
expectation value of the pinwheel density $\langle\rho\rangle$ in
an ensemble of planforms of order $n$, we synthesize planforms (\ref{eq:planforms-specific})
with randomly chosen sets of wavevector directions $l_{j}$ and phases
$\phi_{j}$ in a quadratic region with linear extension $L$ on a
$2048\times2048$ grid choosing an aspect ratio of \textbf{$\Gamma=L/\Lambda=32$}.
In each planform, pinwheel locations are identified within a quadratic
subregion of size $8\Lambda\times8\Lambda$. Realizations are collected
until sufficient precision of average pinwheel densities $\langle\rho\rangle$
is reached, measured by the standard error measure (SEM) $\Delta=s/\sqrt{N}$,
were $s$ is the standard deviation (SD) of densities $\rho$ and
$N$ the number realizations. The afforded precision of $\Delta<0.03$
for $5\leq n\leq14$ and $\Delta<0.01$ for $15\leq n\leq20$ requires
between $100$ and $1000$ realizations per order $n$. At a given
order $n$, we determine the average pinwheel density $\langle\rho\rangle$
and the SD $s$ from the ensemble of realizations. Note that by this
method, both quantities can be evaluated with arbitrary precision
by using sufficient large number of realizations. 

Fig. \ref{cap:planform}a shows the average pinwheel density $\langle\rho\rangle$
of planforms of various orders ranging between $3\le n\le25$. Among
smaller orders $n\le7$, average densities $\langle\rho\rangle$ fluctuate
substantially covering the range $2.5\le\langle\rho\rangle\le3.5$.
However, they are much more confined, between $2.9\le\langle\rho\rangle\le3.2$,
for intermediate orders $8\le n\le15$. For large order $n>15$, the
ensemble average $\langle\rho\rangle$ appears to converge towards
$\pi$ from below. This is consistent with Eq. (\ref{eq:pwd-xi})
showing that already for intermediate orders of $n$ averaging over
$l$ leads to pinwheel densities smaller but close to $\pi$. That
average pinwheel densities $\langle\rho\rangle$ are smaller than
$\pi$ is explained by the upper bound of $\rho_{l}$ in Eq. (\ref{eq:pwd-xi})
implying an upper bound also for its average. That they are, in fact,
not much smaller than $\pi$ is suggested by the {}``relativistic''
form of Eq. (\ref{eq:pwd-xi}). Even moderate anisotropies up to,
e.g., $\xi_{l}=0.3$ result in pinwheel densities $\left\langle \rho_{l}\right\rangle >3$.
Furthermore, as shown in Fig. \ref{cap:planform}b, the variation
$s$ of pinwheel densities $\rho$ in different realizations decreases
drastically with $n$. For large $n$, $s$ became successively smaller
consistent with the limiting value $s^{2}=\mu=0$ derived above. These
results show that even for intermediate interaction ranges, both the
average pinwheel density $\langle\rho\rangle$ of an ensemble of maps
and the pinwheel density $\rho$ of almost every single realization
are close to $\pi$.

\section{Discussion}

We derived signatures of cortical self-organization that can be tested
experimentally. In a model for the self-organization of the system
of orientation preferences in the visual cortex we calculated the
density of pinwheels (topological defects) that the model is predicting.
We find that pinwheel densities close to $\pi$ are robustly selected
if interactions between remote contour detectors are prevalent. Near
criticality ($r\ll1$), in the limit of large intracortical interaction
range $\sigma$, the average pinwheel density converges to the fixed
number $\langle\rho\rangle=\pi$. For intermediate ranges $1\Lambda\lesssim\sigma\lesssim4\Lambda$,
average pinwheel densities $\langle\rho\rangle$ remain smaller but
close to this limit value. For successively larger ranges $\sigma$,
the pinwheel density approaches $\pi$ from below. Moreover, the variation
of pinwheel densities $\rho$ across realizations decreases with interaction
range such that when the interaction range is large, almost every
map exhibits a pinwheel density close to $\pi$. Thus, for a broad
parameter regime of interaction ranges, the model predicts an average
pinwheel density close to $\pi$ even in individual cortical orientation
maps. 

The results presented here are obtained close to the bifurcation point
($r\ll1$). In this regime, the model is well approximated by the
amplitude equations and its behavior is representative for a large
class of models. Thus, the results derived in this regime are robust
against variation of the model. Investigations further away from criticality
at $r>0$ would be instructive, but can at present only be carried
out by numerical methods. While dependencies on the details of the
model are expected to play a role for $r>0$, solutions in this regime
may still be categorized by those obtained for $r\ll1$. 

In the visual cortex, distant neurons can directly and strongly interact
by long-range \textbf{}horizontal connections and by feedback connections
from one visual area to another. \textbf{}In mature cortical circuits,
such connections link neurons that share similar receptive field properties
and represent somewhat displaced locations in visual space\cite{loewel/singer:90,bosking:97}.
In fact, in a wide range of mammals, the range of these connections
is much larger than the spacing of columns\cite{loewel/singer:90,bosking:97,white:01}.
Thus, these species are candidates for testing the value of the average
pinwheel density predicted by our study. 

The approach presented in this study, to our knowledge, for the first
time allows to analytically characterized the statistics of pinwheels
in populations of asymptotically stable states of visual cortical
self-organization. The results obtained are valid for the general
class of Turing-type systems exhibiting the four symmetries Eq. (\ref{eq:transl-sym}-\ref{eq:shift-symm},\ref{eq:permutation-sym})
and dominated by long-ranging interactions when studied close to the
instability threshold. Asymptotically stable states have been studied
in two competing model classes. In one class of models, pinwheels
annihilate dynamically, such that pinwheel density is a time dependent
quantity and no particular finite pinwheel density is intrinsically
selected \cite{wolf:98,koulakov:01,lee:03,xu:05a}. In other models,
a substantial number of pinwheels are preserved in the final state,
but the pinwheels typically crystallize into repetitive arrangements
\cite{koulakov:01,Mayer:2002,lee:03,thomas:04}. These states are
easy to characterize, but are not even qualitatively consistent with
the aperiodic arrangement of visual cortical orientation columns.
The model studied here is distinguished from these competing models
by exhibiting in the long-range interaction dominated regime ($g<1$,
$\sigma>\Lambda$) a large multiplicity of aperiodic stable solutions.
Our study presents a transparent analytical approach for studying
pinwheel density selection in such models. It will be interesting
to generalize this approach to further model classes to comprehensively
clarify whether genuinely different models of visual cortical development
can be distinguished through their pinwheel statistics, as the above
results suggest.

It is natural to ask whether and how the predicted pinwheel density
can be tested experimentally. The currently available published data
is consistent with a pinwheel density around 3 but appears insufficient
to obtain a high precision estimate of this quantity (see e.g. \cite{shmuel:00,muller:00},
for a summary of the earlier literature see \cite{wolf:98}). To obtain
a precise estimate both pattern wavelength and absolute pinwheel number
need to be reliably quantified. Over the past years, image analysis
methods have been devised that enable estimating the pattern wavelength
of orientation maps with a precision in the range of a few percent
(\cite{kaschube:02,kaschube:03}). For instance, using a wavelet method
for the estimation of local pattern wavelength Kaschube and coworkers
showed that genetically related cats often differ in the mean wavelength
of orientation columns by less than 4 \% (\cite{kaschube:02}). As
the count variance of pinwheel number estimates is of order (number
of pinwheels)$^{1/2}$ , estimating the pinwheel density with an accuracy
in the percent range will require data sets encompassing on the order
of 10000 genuine pinwheels. In many model animals, this size of data
set is equivalent to more 100 brain hemispheres that need to be imaged
under consistent experimental conditions. Whereas this may appear
a large animal cohort for any individual study, such large size data
sets will eventually accumulate in many laboratories that use intrinsic
signal optical imaging as a standard technique, enabling to test models
of visual cortical development with quantitative precision.

Acknowledgement: We acknowledge discussions with Theo Geisel, David
Coppola, Siegrid Loewel, Ken Miller and Len White. This work was supported
by grants from the Human Frontiers Science Program and the BMBF to
F.W.

\bibliographystyle{unsrt}
\bibliography{lit_newphys}

\end{document}